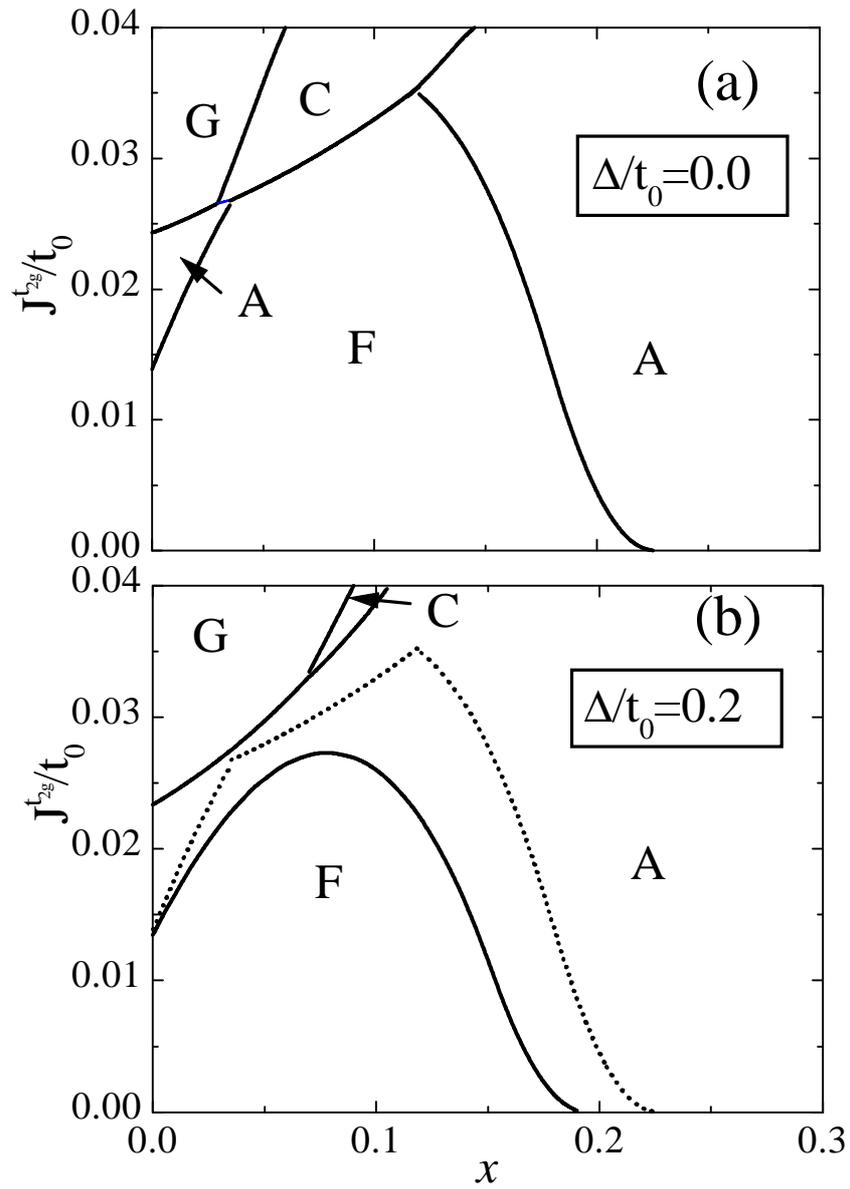

S.Ishihara et al. Fig. 1

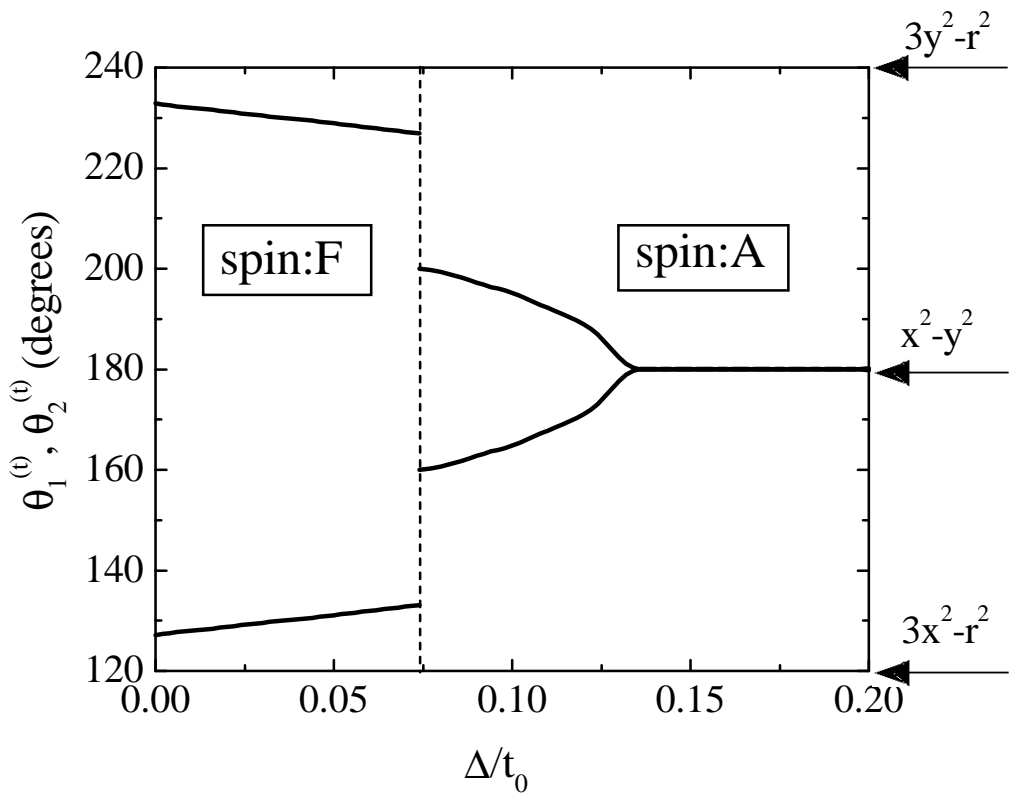

S.Ishihara et al. Fig. 2

# Pressure Effects in Manganites with Layered Perovskite Structure

Sumio Ishihara, Satoshi Okamoto * and Sadamichi Maekawa

*Institute for Materials Research, Tohoku University, Sendai 980-77*

Pressure effects on the charge and spin dynamics in the bilayer manganite compounds $La_{2-2x}Sr_{1+2x}Mn_2O_7$ are studied theoretically by taking into account the orbital degrees of freedom. The orbital degrees are active in the layered crystal structure, and applied hydrostatic pressure stabilizes the $3d_{x^2-y^2}$ orbital in comparison with $3d_{3z^2-r^2}$. The change of the orbital states weakens the interlayer charge and spin couplings, and suppresses the three dimensional ferromagnetic transition. Numerical results, based on an effective Hamiltonian which includes the energy level difference of the orbitals, show that the applied pressure controls the dimensionality of the spin and charge dynamics through changes of the orbital states.

The colossal magnetoresistance (CMR) observed in several manganese oxides has recently attracted considerable attention from both academic and technological viewpoints. In order to improve the technical relevance, several material syntheses and processing methods based on the manganites have been attempted. One of the convincing candidates for the large MR materials is the manganese oxides with layered perovskite structure $La_{1-x}Sr_{1+x}MnO_4$ and $La_{2-2x}Sr_{1+2x}Mn_2O_7$. [1] The layered compounds consist of the single- or double-layered $MnO_2$ planes and the intervening insulating planes stacked along the $c$-axis. Therefore, they are recognized as an intrinsic ferromagnetic metal (FM)-insulator (I)-FM multilayered system.

The unique characteristic of the MR effect observed in $La_{1.4}Sr_{1.6}Mn_2O_7$ is interpreted as the tunneling magnetoresistance. [2] The transport and magnetic properties in this compound are characterized by two temperatures, [2] $T^{ab}_{max}$ and $T^c_{max}$ ($T^{ab}_{max} > T^c_{max}$), where the resistivities in the $ab$-plane ($\rho_{ab}$) and along the $c$-axis ($\rho_c$), respectively, become maximum. In the region above $T^{ab}_{max}$, both $\rho_{ab}$ and $\rho_c$ increase with decreasing temperature. The remarkable feature appears in the intermediate temperature range ($T^{ab}_{max} > T > T^c_{max}$), where $\rho_{ab}$ decreases with decreasing temperature, although $\rho_c$ still exhibits insulating behavior. A large MR value ($\rho_c(H=0)/\rho_c(H=50\text{kOe}) \sim 10000\%$) is observed in this temperature range. It is interpreted that around $T^{ab}_{max}$ the ferromagnetic spin correlation in the $ab$-plane starts growing up. On the other hand, spins are almost randomly oriented along the $c$-axis. Below $T^c_{max}$, the three-dimensional ferromagnetic correlation develops and the resistivities along both directions rapidly decrease.

Recently, the pressure effects on the transport and magnetic properties in $La_{1.4}Sr_{1.6}Mn_2O_7$ are studied by Kimura et al. [3] It is shown that applied hydrostatic pressure up to 11kbar suppresses the steep drops of $\rho_c$ and $\rho_{ab}$ around $T^c_{max}$ and preserves the two-dimensional-like charge and spin dynamics down to low temperatures. In other words, the magnitudes of the interplane magnetic and charge couplings are reduced by the application of pressure. As a result, the applied magnetic field promotes the transport along the $c$-axis and results in a larger MR effect than that obtained under the ambient pressure. This is highly in contrast to the manganites with three-dimensional perovskite structure, where the applied pressure increases the ferromagnetic transition temperature, and it is interpreted as the enhancement of the ferromagnetic double exchange interaction due to the pressure. [4] On the other hand, it is difficult to understand the confinement-like behavior of the spin and charge dynamics observed in the layered compound due to the applied pressure from the conventional double exchange scenario.

In this letter, we study, theoretically, the pressure effects of the transport and magnetic properties in layered perovskite manganites by taking into account the orbital degrees of freedom. It is well known that in manganites with three-dimensional perovskite structure the $e_g$ electron in the $Mn^{3+}$ ion has orbital degrees of freedom, in addition to charge and spin. The orbital plays several important roles in the spin and charge dynamics in manganites, as several authors have pointed out. [5–13] It is undoubted that the orbital degrees are also active in the layered manganites of the present interest. Actually, in recent X-ray diffraction experiments the orbital ordering has been directly observed in $La_{1.5}Sr_{0.5}MnO_4$, [14] which supports the CE-type spin structure [15] known in the cubic compounds. [16, 5] Because the orbital degrees of freedom strongly couples with the lattice, the structural change due to the applied pressure results in a modification of the orbital state. To confirm the change of the orbital state, we calculate the Madelung potential for the $3d_{x^2-y^2}$ and $3d_{3z^2-r^2}$ orbitals, and show that the $3d_{x^2-y^2}$ orbital is relatively stabilized with increasing pressure. By considering the energy-level difference between the two $e_g$ orbitals, we derive the effective Hamiltonian de-scribing the low-energy electronic states, and investigate the spin and orbital phase diagram. The numerical results are satisfactory to explain the suppression of the three-dimensional ferromagnetic state experimentally reported.

---

* On leave from Department of Applied Physics, Nagoya University, Nagoya 464-01

First, let us consider the orbital states under pressure. The crystal structure in $La_{1.2}Sr_{1.8}Mn_2O_7$ under the hydrostatic pressure up to 6.3 kbar has been studied by the neutron diffraction experiment. [17] It is reported that the absolute value of the $c$-axis compressibility is almost three times larger than that in the $ab$-plane in the entire temperature range. In the $MnO_6$ octahedron, the shortening of the Mn-O apical bond is remarkable ($\Delta d/d \sim 1\%$ for 6.3kbar) in comparison with that of the in-plane Mn-O bond ($\sim 0.1\%$) in the ferromagnetic phase. The in-plane Mn-O-Mn bond angle remains unchanged ($\sim 180°$), indicating no tilting of the octahedron. In order to estimate the orbital state under the above structural change, we calculate the Madelung potential which controls the energy levels of the $e_g$ orbitals. The conventional Ewald method is adopted using the structural data at 100K. The Madelung potential for the $3d_{x^2-y^2}$ and $3d_{3z^2-r^2}$ orbitals is given by $V(\boldsymbol{r}_o \pm r_d\boldsymbol{x})$ and $(V(\boldsymbol{r}_o + r_d\boldsymbol{z}) + V(\boldsymbol{r}_o - r_d\boldsymbol{z}))/2$, respectively. Here, $\boldsymbol{r}_o$ is the position of the Mn ion, $r_d(= 0.42\text{Å})$ is the radius where the radial charge density of the Mn 3d orbital is maximum [18] and $\boldsymbol{x}$ and $\boldsymbol{z}$ are the unit vectors along the in-plane and inter-plane Mn-O bonds, respectively. The potential difference between the $3d_{x^2-y^2}$ and $3d_{3z^2-r^2}$ orbitals for the electron is obtained as $(V_{x^2-y^2} - V_{3z^2-r^2}) = 0.0313$eV (0.48kbar), 0.0046eV (3.1kbar) and -0.0007eV (6.3kbar). With increasing pressure, the potential for the $x^2-y^2$ orbital is relatively decreased. Although the crystal structure was measured from 0.48kbar to 6.3kbar, we suppose that the relative energy stabilization for the $x^2-y^2$ orbital amounts to an order of 0.1eV at 11.0kbar where the large pressure effect appears in the resistivity. The above changes of the potential are mainly attributed to the anisotropic change of the Mn-O bond length in the octahedron. We confirm that the hydrostatic pressure acts like an uniaxial pressure applied along the $c$-axis in terms of the stabilization of the $x^2-y^2$ orbital.

Next, we study the spin and charge dynamics by taking into account the change of the orbital state due to the applied pressure. We extend the model Hamiltonian, previously derived for the manganites with cubic structure, [10] to the case of the layered compounds under pressure. For simplicity, we consider the uniaxial three-dimensional crystal structure consisting of Mn ions, instead of the actual layered structure. In this model, the bilayered $MnO_2$ planes are represented by a single plane where the Mn ions form a square lattice. In each Mn site, we introduce two $e_g$ orbitals and a localized spin ($\boldsymbol{S}^{t_{2g}}$) with $S = 3/2$ as the $t_{2g}$ spin. The difference of the energy levels between $a(= 3z^2-r^2)$- and $b(= x^2-y^2)$-orbitals is represented as $\Delta = \varepsilon_a - \varepsilon_b$. The Hund coupling ($J_H$) between $e_g$ and $t_{2g}$ spins and the electron-electron interactions ($U, U', I$) between $e_g$ electrons are introduced. Furthermore, between the nearest neighboring $i$ and $j$ sites, the electron transfer ($t_{ij}^{\gamma\gamma'}$) for the $e_g$ electrons and the antiferromagnetic interaction for the $t_{2g}$ spins are considered. Among the above energy parameters, the electron-electron interactions $U, U'$ give the largest energy scale. Therefore, we derive the effective Hamiltonian to describe the low energy electronic state by the perturbational calculation with respect to (the transfer energy)/(the electron-electron interaction) as follows, $H_{eff} = \widetilde{H}_{e_g} + H_{Hund} + H_{t_{2g}}$. The first term describes the $e_g$ electron system and is given by

$$\widetilde{H}_{e_g} = \Delta \sum_i T_{iz} + \sum_{<ij>\sigma\gamma\gamma'} t_{ij}^{\gamma\gamma'}(\widetilde{d}_{i\gamma\sigma}^\dagger \widetilde{d}_{j\gamma'\sigma} + h.c.)$$
$$+ H_U + H_{U'-I} + H_{U'+I} , \quad (1)$$

where $\widetilde{d}_{i\gamma\sigma}$ is the annihilation operator excluding the double occupancy, and $\boldsymbol{T}_i = \frac{1}{2}\sum_{\alpha\gamma\gamma'} \widetilde{d}_{i\gamma\alpha}^\dagger \boldsymbol{\sigma}_{\gamma\gamma'} \widetilde{d}_{i\gamma'\alpha}$ is the pseudo spin operator for the orbital. $t_{ij}^{\gamma\gamma'}$ is represented as $t_{ij}^{\gamma\gamma'} = \alpha_{\gamma\gamma'} t_{0ij}$ where $\alpha_{\gamma\gamma'}$ is the numerical factor depending on the orbital. [10] The energy-level difference between the two orbitals is represented by $\Delta \sum_i T_{iz}$. The second term and the last three terms in eq.(1) are named as the '$t$-term' and the '$J$-term', respectively, on the analogy of the $t$-$J$ model. Among the three '$J$-terms', $H_{U'-I}$ is the leading term and its explicit formula is given by

$$H_{U'-I} = -\frac{1}{2}\sum_{<ij>} \quad (\frac{3}{4} + \boldsymbol{S}_i \cdot \boldsymbol{S}_j)$$
$$\times \Bigg[ 2\ (t_{ij}^{ab2} + t_{ij}^{ba2})(f_- n_{i+}n_{j+} + f_+ n_{i-}n_{j-})$$
$$+ 2\ f_0(t_{ij}^{aa2} + t_{ij}^{bb2})(n_{i+}n_{j-} + n_{i-}n_{j+})$$
$$- t_{ij}^{ab}t_{ij}^{ba} 2(f_+ + f_-)(T_{i+}T_{j+} + T_{i-}T_{j-})$$
$$- t_{ij}^{aa}t_{ij}^{bb} 4f_0(T_{i+}T_{j-} + T_{i-}T_{j+})$$
$$+ (\ t_{ij}^{ab}t_{ij}^{bb} - t_{ij}^{ba}t_{ij}^{aa})(T_{i+} + T_{i-})$$
$$\times \Big((f_0 + f_-)n_{j+} - (f_0 + f_+)n_{j-}\Big)$$
$$+ (\ t_{ij}^{ba}t_{ij}^{bb} - t_{ij}^{ab}t_{ij}^{aa})(T_{j+} + T_{j-})$$
$$\times \Big((f_0 + f_-)n_{i+} - (f_0 + f_+)n_{i-}\Big)\Bigg], (2)$$

with $\boldsymbol{S}_i = \frac{1}{2}\sum_{\alpha\beta\gamma} \widetilde{d}_{i\gamma\alpha}^\dagger \boldsymbol{\sigma}_{\alpha\beta} \widetilde{d}_{i\gamma\beta}$ and $n_{i\pm} = \frac{1}{2}n_i \pm T_{iz}$. $f_0 = \frac{1}{U'-I}$ and $f_\pm = \frac{1}{U'-I\pm\Delta}$ which are the inverse of the intermediate energy including the energy difference between $a$ and $b$ orbitals. The second and third terms in $H_{eff}$ are shown by

$$H_{Hund} + H_{t_{2g}} = J_H \sum_i \boldsymbol{S}_i \cdot \boldsymbol{S}_i^{t_{2g}} + \sum_{<ij>} J_{ij}^{t_{2g}} \boldsymbol{S}_i^{t_{2g}} \cdot \boldsymbol{S}_j^{t_{2g}} ,$$
$$(3)$$

where $J^{t_{2g}}$ is the antiferromagnetic superexchange interaction between the nearest neighbor $t_{2g}$ spins. The uniaxial crystal structure is reflected in the anisotropy of the transfer intensity $t_{ij}^{\gamma\gamma'}$ and the superexchange interaction $J_{ij}^{t_{2g}}$, and their ratio is represented as $R = t_{0i,i+x}/t_{0i,i+z} = \sqrt{(J_{i,i+x}^{t_{2g}}/J_{i,i+z}^{t_{2g}})}$. For simplicity, $t_{0i,i+x}$ and $J_{i,i+x}^{t_{2g}}$ are represented as $t_0$ and $J^{t_{2g}}$, respectively. The effects of the applied hydrostatic pressure is described by the change of the energy level difference $\Delta$. When we assume as $\Delta = 0$ and $R = 1$, the Hamiltonian is reduced to the isotropic one. [10] In the following numerical calculation, we use the model which is expanded with respect to $\Delta/U$ up to the first order. Because the $\Delta$ dependent part of the 'J-term' shows $O(\Delta t^2/U^2)$, the main contributions of the level difference comes from the first term in eq. (1). When the transfer matrices are assumed to be $t_{ij}^{ab} = \delta_{ab} t_{0ij}$ in eq. (1) at $x = 0$, we obtain the effective Hamiltonian, where the orbital space is isotropic, [19,20] plus $\Delta \sum_i T_{iz}$. In this model, the ground state changes from the well-known ferromagnetic spin ordering (spin:F) accompanied with the antiferromagnetic orbital ordering (orbital:AF) to the spin:AF with the orbital:F at $\Delta \sim 3I(t^2/U')^2$. In other words, the spin ferromagnetic state becomes unstable by lifting the orbital degeneracy.

Based on the above Hamiltonian, we calculate the spin and orbital phase diagram in the mean field approximation. We consider four types of the spin and orbital orderings in the cubic eight Mn unit; ferromagnetic (F) ordering and layer-type (A), rod-type (C) and NaCl-type (G) antiferromagnetic(AF) orderings. In the 'J-term', $\langle S_{iz} \rangle$ and $\langle n_i \rangle$ are adopted as the order parameters. Because of the anisotropy in the orbital space, we introduce the rotating frame in the 'J-term' where the orbital state is described by the angles $(\theta_1^{(t)}, \theta_2^{(t)})$ for each orbital sublattice as $|\theta_i^{(t)}\rangle = \cos(\theta_i^{(t)}/2)|3z^2 - r^2\rangle - \sin(\theta_i^{(t)}/2)|x^2 - y^2\rangle$. Then, we adopt $\langle \tilde{T}_{iz} \rangle (= \cos \theta_i^{(t)} \langle T_{iz} \rangle + \sin \theta_i^{(t)} \langle T_{ix} \rangle)$ as the order parameter for the orbital. By the same token, the rotating frame is adopted in the 't-term' by introducing the unitary matrices in the spin and orbital spaces. $\tilde{d}_{i\sigma\gamma}$ is transformed as $\tilde{d}_{i\sigma\gamma} = h_i^\dagger z_{i\sigma}^{(s)} z_{i\gamma}^{(t)}$ where $h_i$ is a fermion operator describing the hole motion, and $z_{i\sigma}^{(s)}$ and $z_{i\gamma}^{(t)}$ are the elements of the unitary matrix in the spin and orbital frame, respectively. Then, 't-term' is rewritten as $-\sum_{<ij>} h_i^\dagger h_j (\sum_\sigma z_{i\sigma}^{(s)*} z_{j\sigma}^{(s)})(\sum_{\gamma\gamma'} z_{i\gamma}^{(t)*} t_{ij}^{\gamma\gamma'} z_{j\gamma'}^{(t)})$, [11]. In order to qualitatively discuss the doping dependence of the phase diagram in the low doping region, we adopt the mean field approximation and replace $\langle h_i^\dagger h_j \rangle$ by the carrier number $x$. $\langle z_{i\sigma}^{(s)*} z_{j\sigma}^{(s)} \rangle$ and $\langle z_{i\gamma}^{(t)*} t_{ij}^{\gamma\gamma'} z_{j\gamma'}^{(t)} \rangle$ are represented by the angles in the spin and orbital spaces, respectively. When we neglect the azimuthal angle in the spin space, $\sum_\sigma \langle z_{i\sigma}^{(s)*} z_{j\sigma}^{(s)} \rangle$ reproduces the form as $\cos((\theta_i^{(s)} - \theta_j^{(s)})/2)$ derived by Anderson and Hasegawa. [21] The spin and orbital phase diagram at $T = 0$ as a function of $x$ and $J^{t_{2g}}$ is obtained by minimizing the energy.

In Fig. 1, we present the numerical results of the spin and orbital phase diagram. In the numerical calculation, we only consider the leading term $H_{U'-I}$ among the three 'J-terms'. The parameter values are chosen as $U' = 3.0t_0$, $I = 1.0t_0$ and $R = 2.0$. First, we focus on the $\Delta = 0.0$ case (Fig. 1(a)). It is noted that in the case of $R = 1$, the present calculation well reproduces the results obtained by the exact diagonalization method based on the effective Hamiltonian at $x = 0$ [12] and the Hartree-Fock calculation in the region of finite $x$. [13] Therefore, the lack of the spin:C phase at $x = 0.0$ between the spin:A and spin:G phases is attributed to the anisotropy in $t_{ij}^{\gamma\gamma'}$ and $J_{ij}^{t_{2g}}$. With increasing the carrier concentration from $x = 0.0$, the region of the ferromagnetic phase gradually grows up, and becomes maximum around $x = 0.1$. With further increasing the concentration, this phase disappears at $x \cong 0.2$ and the A-type AF phase becomes stable in the large parameter region. The stabilization of the A-type AF in the region of higher $x$ originates from the $x^2 - y^2$ orbital ordering which is most favorable to the gain of the kinetic energy of the doped carriers, as indicated by the Hartree-Fock calculation. [13] Although the superexchange and double exchange interactions contribute to the ferromagnetic phase which appears around $0.0 < x < 0.2$, the former is mainly dominant. The orbital state in the ferromagnetic phase is gradually changed from $(\theta_1^{(t)} = 90°, \theta_2^{(t)} = 270°)$ at around $x = 0.0$ to $(\theta_1^{(t)} \cong 120°, \theta_2^{(t)} \cong 240°)$ around $x = 0.2$. When $\Delta$ is introduced (Fig. 1(b)), remarkable changes appear around the boundary between the spin:F and spin:A phases, that is, the large region of the ferromagnetic phase is replaced by the spin:A phase. Around the boundary, spin:F accompanied with $(\theta_1^{(t)} \cong 120°, \theta_2^{(t)} \cong 240°)$ energetically competes with spin:A with the $x^2 - y^2$-type orbital ordering. Therefore, the small value of $\Delta$ relatively stabilizes the spin:A phase. The suppression of the three-dimensional ferromagnetic phase originates primarily due to the following two reasons: i) the magnitude of the ferromagnetic superexchange interaction is reduced by lifting the orbital degeneracy, and ii) both super and double exchange interactions in the $c$-direction are prevented in the $x^2 - y^2$ orbital ordering. On the other hand, the spin:F phase around $x = 0.0$ is still robust in the case of $\Delta = 0.1t_0$, because this phase does not adjoin the other spin phase accompanied with the $x^2 - y^2$-type ordering. In Fig. 2, we present the $\Delta$ dependence of the angles $(\theta_1^{(t)}, \theta_2^{(t)})$ in the orbital space at $x = 0.16$ and $J_{i,i+x}^{t_{2g}} = 0.017t_0$. As the magnitude of $\Delta$ is increased, the pseudo-spins gradually cant from $(\theta_1^{(t)} \cong 120°, \theta_2^{(t)} \cong 240°)$ toward $(\theta_1^{(t)} = \theta_2^{(t)} = 180°)$ ($x^2 - y^2$ orbital) within the spin:F phase, and finally the spin:A appears at $\Delta = 0.075t_0$. It is clearly shown that the applied pressure tends to align the orbital pseudo-spin ferromagnetically, like an applied magnetic field on the spin space.

It is shown by the numerical calculations that the spin:F phase near the boundary is replaced by the spin:A phase by introducing $\Delta$. In this phase, the electron transfer is prohibited in the $c$-direction, and the interlayer spin coupling is dominated only by the weak antiferromagnetic interaction $J_{i,i+z}^{t_{2g}}$ derived by the electron transfer in the Mn 3d-O 2p $\pi$ bond. As a result, both spin and charge couplings are considerably weak along the $c$-axis in comparison with those in the $ab$-plane. Also in the state where the ferromagnetic phase remains under the applied pressure, the weight of the $x^2 - y^2$ orbital becomes dominant as shown in Fig. 2, and the interlayer coupling is weakened. Therefore, the above states are regarded as the two-dimensional ferromagnetic state with the weak interplane spin coupling, corresponding to the state observed in $La_{1.4}Sr_{1.6}Mn_2O_7$ under pres-sure. Although our calculation is limited in the mean field phase diagram, where the qualitative discussion is limited in the low concentration region and the spin and orbital fluctuations are neglected, the obtained result is sufficient to explain the confinement of the spin and charge dynamics within the $ab$-plane by the application of pressure. In our calculation, the actual layered structure is replaced by the anisotropic three-dimensional one, as explained above. However, the ferromagnetic coupling within the bilayers is expected to remain under the applied pressure as the experimental results show; the length of the Mn-O-Mn bond linkage between two $MnO_2$ planes shows weaker changes in comparison with that of the Mn-O apical bond. Also, the present results are consistent with the temperature dependence of crystal structure in $La_{1.2}Sr_{1.8}Mn_2O_7$, where the Mn-O apical bond length is sharply increased below the ferromagnetic transition temperature. [22] The applied pressure suppresses an elongation of the apical bond and stabilizes the $x^2 - y^2$ orbital. As a result, the ferromagnetic transition is restrained. It is noted that by application of uniaxial pressure along the $c$-axis, the two-dimensional properties are expected to be more significant in comparison with those in the case of hydrostatic pressure.

In summary, we investigate the pressure effects on the spin and orbital states in the double layered perovskite manganites. Because of the anisotropic change of the Mn-O bond length, the change of the Madelung potential by the application of pressure stabilizes the $3d_{x^2-y^2}$ orbital in comparison with the $3d_{3z^2-r^2}$ orbital. The spin and charge dynamics are restricted within the $ab$-plane under the $3d_{x^2-y^2}$ orbital state. As a result, the three-dimensional ferromagnetic state is replaced by the two-dimensional one accompanied with the weak interplane spin coupling. The applied pressure in the layered compound is recognized as a tool to control the dimensionality of the spin and charge dynamics through the changes of the orbital.

ACKNOWLEDGMENTS

The authors would like to thank Y.Tokura, T.Kimura, N.Nagaosa, W.Koshibae and J.Inoue for their valuable discussions. The authors also thank T.Tohyama and Y.Mizuno for their assistance in the calculation of the Madelung potential. This work was supported by Priority Areas Grants from the Ministry of Education, Science and Culture of Japan, and NEDO of Japan. Part of the numerical calculation was performed in the HITACS-3800/380 supercomputing facilities in IMR, Tohoku Univ..

*Note added in proof* — The present scenario also explains the experimental results that a magnetic field compensates the effects of hydrostatic pressure shown in Ref. 3. The recent structural studies by D.N.Argyriou et al. [Phys. Rev. B **55** (1997) R11965] have shown that the Mn-O apical bond length is increased by $\sim 1\%$ in 0.6T. This large elongation of the Mn-O apical bond stabilizes the $3z^2 - r^2$ orbital. As a result, the system becomes insensitive to the pressure.

Figure captions

Fig. 1: The spin phase diagram with and without the energy level difference $\Delta$ between two $e_g$ orbitals. The parameter values are chosen as $U' = 3.0t_0$, $I = 1.0t_0$ and $R = 2.0$. The dotted line in Fig. 1(b) shows the phase boundary between the spin:F phase and the other spin phases for $\Delta = 0.0$ case.

Fig. 2: The angles $(\theta_1^{(t)}, \theta_2^{(t)})$ in the orbital space as a function of $\Delta$. The parameter values are chosen as $U' = 3.0t_0$, $I = 1.0t_0$, $J^{t_{2g}} = 0.017t_0$, $x = 0.16$ and $R = 2.0$. The angles 120°, 180° and 240° correspond to the $d_{3x^2-r^2}$, $d_{x^2-y^2}$ and $d_{3y^2-r^2}$ orbitals, respectively.